\documentclass[aps,preprint,nofootinbib]{revtex4}
\usepackage{amsfonts}
\usepackage{amsmath}
\usepackage{amssymb}
\usepackage{graphicx}

\input{tcilatex}
\begin{document}

\title{Algebraic analysis of a model of two-dimensional gravity}
\author{A.M. Frolov}
\affiliation{Department of Chemistry, University of Western Ontario, London, Ontario,
Canada}
\email{afrolov@uwo.ca}
\author{N. Kiriushcheva}
\affiliation{Department of Applied Mathematics, University of Western Ontario, London,
Ontario, Canada}
\email{nkiriush@uwo.ca}
\author{S.V. Kuzmin}
\affiliation{Faculty of Arts and Social Science, Huron University College and Department
of Applied Mathematics, University of Western Ontario, London, Ontario,
Canada}
\email{skuzmin@uwo.ca}
\date{\today}

\begin{abstract}
An algebraic analysis of the Hamiltonian formulation of the model
two-dimensional gravity\ is performed. The crucial fact is an exact
coincidence of the Poisson brackets algebra of the secondary constraints of
this Hamiltonian formulation with the SO(2,1)-algebra. The eigenvectors of
the canonical Hamiltonian $H_{c}$ are obtained and explicitly written in
closed form.

PACS number(s): 11.10.Ef and 11.30.Cp (also 02.20.Sv)
\end{abstract}

\maketitle

\section{Introduction}

In this letter we consider the model which arises from the Einstein-Hilbert
(EH) action \cite{LLF, Carmeli}

\begin{equation}
S_{D}=\int d^{D}x\sqrt{\left( -1\right) ^{D-1}g}g^{\mu \nu }R_{\mu \nu
}\left( g\right)  \label{eqn1}
\end{equation}%
where $R_{\mu \nu }=$ $\Gamma _{\mu \nu ,\lambda }^{\lambda }-\Gamma _{\mu
\lambda ,\nu }^{\lambda }+\Gamma _{\sigma \lambda }^{\lambda }\Gamma _{\mu
\nu }^{\sigma }-\Gamma _{\sigma \mu }^{\lambda }\Gamma _{\lambda \nu
}^{\sigma }$ is the Ricci tensor, $\Gamma _{\mu \nu }^{\lambda }=\frac{1}{2}%
g^{\lambda \sigma }\left( g_{\mu \sigma ,\nu }+g_{\nu \sigma ,\mu }-g_{\mu
\nu ,\sigma }\right) $ is the affine connection and the action (\ref{eqn1})
is written in terms of a metric tensor $g_{\mu \nu }$ and its second and
first derivatives (comma \textquotedblleft ,\textquotedblright\ indicates
the differentiation and $D$ is a spacetime dimension). This is a
\textquotedblleft second-order\textquotedblright\ formalism. If we treat $%
g_{\alpha \beta }$ and $\Gamma _{\mu \nu }^{\lambda }$ as independent
variables, then we have a \textquotedblleft first-order\textquotedblright\
formulation

\begin{equation}
S_{D}=\int d^{D}x\sqrt{\left( -1\right) ^{D-1}g}g^{\mu \nu }\left( \Gamma
_{\mu \nu ,\lambda }^{\lambda }-\Gamma _{\mu \lambda ,\nu }^{\lambda
}+\Gamma _{\sigma \lambda }^{\lambda }\Gamma _{\mu \nu }^{\sigma }-\Gamma
_{\sigma \mu }^{\lambda }\Gamma _{\lambda \nu }^{\sigma }\right)
\label{eqn2}
\end{equation}%
which was originally introduced by Einstein \cite{Einstein} (not by
Palatini, as it is generally believed \cite{Palatini}). By solving equations
of motion for $\Gamma _{\mu \nu }^{\lambda }$ in terms of $g_{\mu \nu }$ and
substituting the solutions to (\ref{eqn2}), it is easy to show the
equivalence of these second- (\ref{eqn1}) and first-order (\ref{eqn2})
formulations of EH action for the dimensions of spacetime $D$ higher than
two ($D>2$) \cite{Ann1}.

In $D=2$ the field equations cannot be solved for $\Gamma _{\mu \nu
}^{\lambda }$ in terms of $g_{\mu \nu }$ \cite{Lingstrom, Mann, Ann1}, which
is why equation (\ref{eqn2}) does not provide an equivalent first-order
formulation of EH action in $2D$. For the Hamiltonian treatment of real
two-dimensional gravity in second-order form see \cite{OnHam, GKK}. Although
the action (\ref{eqn2}) does not reproduce the real $2D$ gravity, it can
nevertheless be treated as a model which we call \textquotedblleft
two-dimensional gravity\textquotedblright\ (or 2DG for short), remembering
that it is not equivalent to the second-order EH action when $D=2$. This
model is no worse than any other two-dimensional model arising from
modifications of the EH action (e.g., dilaton $2D$ gravity,
\textquotedblleft string-inspired\textquotedblright\ $2D$ gravity, etc.).

In addition, and what is more important, this model can provide a deep
insight into the first-order, affine-metric, formulation of the EH action in
higher dimensions \cite{Ann1, KKM2}. First of all, the action (\ref{eqn2})
is indeed equivalent to the original second-order EH action (\ref{eqn1}) in
dimensions $D>2$ \cite{Ann1}. Second, the structure of constraints in the
2DG model is much closer to the higher dimensional first-order gravity (\ref%
{eqn2}) (see \cite{Ann1, KKM2, G, GM}) than the structure of constraints of
the real $2D$ gravity (see \cite{OnHam, GKK}). As it was shown in \cite%
{OnHam}, the Hamiltonian formulation of the second-order EH action (or real $%
2D$ gravity) in two dimensions leads to three primary first class
constraints which generate the gauge transformations consistent with zero
degrees of freedom and triviality of the Einstein equations in $2D$. If the
constraints structure of the real $2D$ gravity imitated those of the higher
dimensional first-order EH action, then there would be at least two primary
and two secondary first class constraints.\footnote{%
In \cite{Ann1, KKM2} we showed that in the Hamiltonian formulation of the
first-order EH action in dimensions $D>2$ tertiary constraints should also
appear which is consistent with counting of degrees of freedom. The explicit
form of tertiary constraints as well as the closure of the Dirac procedure
was demonstrated in \cite{G, GM} for all dimensions $D>2$.} This would
produce in $2D$ minus one degree of freedom \cite{Mart} meaning that the
system (in such a formulation) is overconstrainted and non-physical. In 2DG
model three primary and three secondary first class constraints appear which
is also consistent with being zero degrees of freedom, but in contrast with
the real $2D$ gravity \cite{OnHam, GKK}, two of the secondary constraints of
the 2DG model, (\ref{c2}, \ref{c3}), as well as the Poisson brackets among
them (\ref{pb1}), are exactly the same as in higher dimensional first-order
EH action if we replace in (\ref{c2}, \ref{c3}) the index \textquotedblleft
1\textquotedblright\ by \textquotedblleft $k$\textquotedblright\ or
\textquotedblleft $n$\textquotedblright\ ($k,n=1,2,...,D-1$). For details
see \cite{Ann1, KKM2, G, GM}.

The canonical analysis of the two-dimensional gravity model can be found in 
\cite{KKM1, KKM2, Ann1, Gerry}. We will briefly outline here the Hamiltonian
formulation of this model. The Lagrangian density is

\begin{equation}
L_{2}=h^{\mu \nu }\left( \Gamma _{\mu \nu ,\lambda }^{\lambda }-\Gamma _{\mu
\lambda ,\nu }^{\lambda }+\Gamma _{\sigma \lambda }^{\lambda }\Gamma _{\mu
\nu }^{\sigma }-\Gamma _{\sigma \mu }^{\lambda }\Gamma _{\lambda \nu
}^{\sigma }\right)  \label{eqn3}
\end{equation}%
where $h^{\mu \nu }$ is the metric density: $h^{\mu \nu }=$ $\sqrt{-g}g^{\mu
\nu }$ ($\mu ,\nu =0,1$ are the spacetime indices). Note that in $2D$ we
cannot express $g^{\mu \nu }$ in terms of $h^{\mu \nu }$ because $h=\det
\left( h^{\mu \nu }\right) =-\left( -g\right) ^{\frac{D-2}{2}}$, so that in
two dimensions $h=-1$; however, the metric tensor appears in the Lagrangian
in the combination $\sqrt{-g}g^{\mu \nu }$.

The analysis is simplified if we use instead of $\Gamma _{\mu \nu }^{\lambda
}$ the linear combination

\begin{equation}
\xi _{\alpha \beta }^{\lambda }=\Gamma _{\alpha \beta }^{\lambda }-\frac{1}{2%
}(\delta _{\alpha }^{\lambda }\Gamma _{\beta \sigma }^{\sigma }+\delta
_{\beta }^{\lambda }\Gamma _{\alpha \sigma }^{\sigma }).  \label{eqn4}
\end{equation}

This covariant change of variables ($\Gamma _{\alpha \beta }^{\lambda
}\rightarrow $ $\xi _{\alpha \beta }^{\lambda }$) provides an alternative
first-order formulation of the EH action in dimensions $D>2$ and for $D=2$
it gives the alternative two-dimensional gravity model with the Lagrangian
density

\begin{equation}
L_{2}=h^{\mu \nu }\left( \xi _{\mu \nu ,\lambda }^{\lambda }-\xi _{\mu
\sigma }^{\lambda }\xi _{\nu \lambda }^{\sigma }+\xi _{\mu \lambda
}^{\lambda }\xi _{\nu \sigma }^{\sigma }\right) .  \label{eqn5}
\end{equation}

We consider (\ref{eqn5}) as a model, treating $h^{\mu \nu }$ and $\xi
_{\alpha \beta }^{\lambda }$ as independent variables and the Lagrangian
density $\widetilde{L}=L-\left( h^{\alpha \beta }\xi _{\alpha \beta
}^{\lambda }\right) _{,\lambda }$. Using this Lagrangian density, $%
\widetilde{L}$, as a starting point of the Hamiltonian formulation allows
completely avoid any integration as constraints (see below) follow directly
from the Hamiltonian, contrary to the usual case when an additional spatial
integration is often needed to single out the common field that appears in
front of a constraint. The general discussion of the role of boundary terms
can be found in \cite{RT}. 

Introducing momenta $\pi _{\alpha \beta }$ and $\Pi _{\lambda }^{\mu \nu }$
conjugate to the variables $h^{\alpha \beta }$ and $\xi _{\mu \nu }^{\lambda
}$, respectively, with the fundamental Poisson brackets (PB) among them $%
[h^{\alpha \beta },\pi _{\mu \nu }]=\Delta _{\mu \nu }^{\alpha \beta }=\frac{%
1}{2}(\delta _{\mu }^{\alpha }\delta _{\nu }^{\beta }+\delta _{\mu }^{\beta
}\delta _{\nu }^{\alpha })$ and $[\xi _{\alpha \beta }^{\lambda },\Pi
_{\sigma }^{\mu \nu }]=\delta _{\sigma }^{\lambda }\Delta _{\mu \nu
}^{\alpha \beta }$, we obtain the primary constraints 
\begin{equation}
\Phi _{\alpha \beta }=\pi _{\alpha \beta }+\xi _{\alpha \beta }^{0}\approx
0,\quad \Phi _{\lambda }^{\mu \nu }=\Pi _{\lambda }^{\mu \nu }\approx 0.
\label{eqn6}
\end{equation}

The total Hamiltonian density is defined as 
\begin{equation}
H_{T}=H_{c}+h_{,0}^{\alpha \beta }\Phi _{\alpha \beta }+\xi _{\mu \nu
,0}^{\lambda }\Phi _{\lambda }^{\mu \nu }
\end{equation}%
where 
\begin{equation}
H_{c}=h_{,0}^{\alpha \beta }\pi _{\alpha \beta }+\xi _{\mu \nu ,0}^{\lambda
}\Pi _{\lambda }^{\mu \nu }-\widetilde{L}_{2}.
\end{equation}

There is the second class subset of constraints among those in equation (\ref%
{eqn6}) which is of a special form \cite{GT}. Because these constraints are
of a special form, we can eliminate the canonical pair of variables

\begin{equation}
\Pi _{0}^{\mu \nu }=0,\quad \xi _{\alpha \beta }^{0}=-\pi _{\alpha \beta }
\label{eqn7}
\end{equation}%
from the total Hamiltonian and constraints. We then obtain the reduced total
Hamiltonian density

\begin{equation}
H_{T}=H_{c}+\xi _{\mu \nu ,0}^{1}\Pi _{1}^{\mu \nu }.
\end{equation}

The conservation in time of the primary constraints $\Phi _{1}^{\mu \nu
}=\Pi_{1}^{\mu \nu }$ leads to the secondary constraints:

\begin{equation}
\chi _{1}^{11}=-(h_{,1}^{11}+2h^{11}\pi _{01}+2h^{01}\pi _{00}),  \label{c1}
\end{equation}

\begin{equation}
\chi _{1}^{01}=-(h_{,1}^{01}-h^{11}\pi _{11}+h^{00}\pi _{00}),  \label{c2}
\end{equation}

\begin{equation}
\chi _{1}^{00}=-(h_{,1}^{00}-2h^{01}\pi _{11}-2h^{00}\pi _{01}).  \label{c3}
\end{equation}

The canonical Hamiltonian density $H_{c}$

\begin{equation}
H_{c}=-\xi _{11}^{1}\chi _{1}^{11}-2\xi _{01}^{1}\chi _{1}^{01}-\xi
_{00}^{1}\chi _{1}^{00}  \label{hc}
\end{equation}%
is just a linear combination of the secondary first class constraints (\ref%
{hc}). The secondary constraints $\chi _{1}^{\alpha \beta }$ have zero PB
with the primary constraints $\Phi _{1}^{\alpha \beta }$ and among
themselves have the following PBs

\begin{equation}
\lbrack \chi _{1}^{01}(x,t),\chi _{1}^{00}(y,t)]=\chi _{1}^{00}(x,t)\delta
(x-y),  \label{pb1}
\end{equation}

\begin{equation}
\lbrack \chi _{1}^{01}(x,t),\chi _{1}^{11}(y,t)]=-\chi _{1}^{11}(x,t)\delta
(x-y),  \label{pb2}
\end{equation}

\begin{equation}
\lbrack \chi _{1}^{11}(x,t),\chi _{1}^{00}(y,t)]=2\chi _{1}^{01}(x,t)\delta
(x-y).  \label{algb}
\end{equation}

The complete analysis of this model can be found in \cite{Ann1}. In next
sections we present the analysis of the model based on local coincidence of
the constraint algebra of (\ref{algb}) with the Lie algebra of SO(2,1).

\section{Algebraic analysis of the model 2DG}

Introducing the three operators $K_{-}=\chi _{1}^{11},K_{0}=\chi _{1}^{01}$
and $K_{+}=\chi _{1}^{00}$, equation (\ref{algb}) takes the form 
\begin{equation}
\lbrack
K_{0},K_{+}]=K_{+}\;\;\;,\;\;\;[K_{0},K_{-}]=-K_{-}\;\;\;,\;\;%
\;[K_{-},K_{+}]=2K_{0}.  \label{algb1}
\end{equation}%
Equations (\ref{algb1}) coincide with the commutation relations for the
three generators of the SO(2,1)-algebra (see, e.g., Eq. (5.14) in \cite{Perl}%
). This coincidence of the PB between the secondary constraints in the 2DG
model and the generators of the SO(2,1)-algebra means that there is a
uniform relation between the corresponding representations of these two
algebras. The Hamiltonian $H_{c}$, equation (\ref{hc}), can now be written
in the form 
\begin{equation}
H_{c}=-2\xi _{01}^{1}K_{0}-\xi _{11}^{1}K_{-}-\xi _{00}^{1}K_{+}=-2\xi
_{01}^{1}K_{0}-\imath (\xi _{11}^{1}-\xi _{00}^{1})K_{1}+(\xi _{11}^{1}+\xi
_{00}^{1})K_{2},  \label{hc1}
\end{equation}%
i.e. as a linear combination of the three generators of the SO(2,1) algebra.
In this equation we take 
\begin{equation}
K_{1}=\frac{\imath }{2}(K_{-}-K_{+})\;\;\;,\;\;\;K_{0}=K_{0}\;\;\;,\;\;%
\;K_{2}=-\frac{1}{2}(K_{+}+K_{-}),  \label{K}
\end{equation}%
so that $K_{\pm }=\pm \imath (K_{1}\pm \imath K_{2})$. The coefficients in
this linear form, (\ref{hc1}), are some $\xi -$numbers which ensure the
correct relation with General Relativity (GR) (see below). Note that
equations (\ref{hc}) and (\ref{hc1}) are the simplest linear forms which are
acceptable for the $H_{c}$ Hamiltonian in General Relativity.

By studying the relation between the algebra of secondary constraints of
this 2DG model and the SO(2,1)-algebra we can come to some conclusions about
properties and spectra of the Hamiltonian of 2DG. Furthermore, for any
operator represented as a linear combination of generators of the
SO(2,1)-algebra one can apply a simple procedure which allows one to
determine all eigenvalues and the corresponding eigenvectors. In the present
case this procedure is based on the use of coherent states constructed from
the SO(2,1)-algebra \cite{Perl}.

The eigenvectors of $H_{c}$ are the classical eigenstates. However, they are
closely related to the corresponding quantum states. Indeed, for any
dynamical system with a classical analogue, a state for which quantum
description is valid is represented in quantum mechanics by a wave packet 
\cite{Dirac2}. In the Schr\"{o}dinger representation such a wave function is
of the form $\Psi (q,t)=A(q,t)\cdot \exp (\imath \frac{S(q,t)}{\hbar })$,
where $A$ and $S$ are the amplitude and phase of the total wave function $%
\Psi $. It can be shown \cite{Dirac2} that the phase function $S(q,t)$
satisfies the following equation (to the lowest order in $\hbar $) 
\begin{equation}
\frac{\partial S}{\partial t}=-H_{c}\left( q,\frac{\partial S}{\partial q}%
\right)
\end{equation}%
which is known as the Hamilton-Jacobi equation (compare with equation (\ref%
{e5}) below). Note that this equation involves the classical Hamiltonian $%
H_{c}$, in which all momenta are replaced by the corresponding partial
derivatives of the Jacobi function $S$.

The eigenvectors of $H_{c}$ can be used to obtain further information about
the Hamiltonian formulation of the 2DG model. Furthermore, using the
conclusions drawn from the 2DG model we can predict some useful properties
of the Hamiltonian $H_{c}$ in four-dimensional and $N-$dimensional GR. This
is the main goal of our work.

\section{Self-adjoint irreducible representations of the SO(2,1)-algebra}

First of all, let us describe the self-adjoint irreducible representations
of the SO(2,1)-algebra \cite{Gelf}. By using the operators $K_{-},K_{+}$ and 
$K_{0}$ we can construct the Casimir operator $\hat{C}_{2}$ 
\begin{equation}
\hat{C}_{2}=K_{0}^{2}-\frac{1}{2}%
(K_{+}K_{-}+K_{-}K_{+})=K_{0}^{2}-K_{0}-K_{+}K_{-}.
\end{equation}%
This operator commutes with the $K_{-},K_{+},K_{0}$. Moreover, $\hat{C}_{2}$
commutes with an arbitrary function of these three operators. In particular,
it commutes with the $H_{c}$ operator defined above. As follows from Schur's
lemma, the operator $\hat{C}_{2}$ is diagonal, i.e., $\hat{C}_{2}=\lambda 
\hat{I}$, where $\hat{I}$ is the unit operator. In the case of
SO(2,1)-algebra the numerical constant $\lambda $ is designated as $k(k-1)$,
where $k$ is some number (see below).

All self-adjoint and irreducible representations of the SO(2,1)-algebra can
be constructed with the use of $\hat{C}_{2}$ and $K_{0}$ operators. In
general, for the SO(2,1)-algebra one finds the two discrete and two
continuous series of representations (see, e.g., \cite{Gelf}). There are the
positive and negative discrete series, and principal and supplementary
series of continuous representations. First, consider the positive discrete
series. Analysis of the negative discrete series is almost identical and
essentially based on the use of the same derivation. In the case of positive
discrete series the two following conditions must be obeyed for each of the $%
\mid k,m\rangle $ basis vectors 
\begin{equation}
\hat{C}_{2}\mid k,m\rangle =k(k-1)\mid k,m\rangle \;\;\;,\;\;\;K_{0}\mid
k,m\rangle =m\mid k,m\rangle,  \label{basic}
\end{equation}%
where $k$ is semi-integer: $k=1,\frac{3}{2},2,\frac{5}{2},\ldots $, while $%
m=k,k+1,\ldots ,k+n,\ldots $ ($n$ is a nonnegative integer). In other words,
we need to determine these vectors from the following eigenvalue equations 
\begin{equation}
\lbrack (\chi _{1}^{01})^{2}-\chi _{1}^{01}-\chi _{1}^{11}\chi
_{1}^{00}]\mid k,m\rangle =k(k-1)\mid k,m\rangle \;\;\;,\;\;\;\chi
_{1}^{01}\mid k,m\rangle =m\mid k,m\rangle.  \label{basic1}
\end{equation}%
For continuous series of representations the last two equations are
traditionally written in the form 
\begin{equation}
\lbrack (\chi _{1}^{01})^{2}-\chi _{1}^{01}-\chi _{1}^{11}\chi
_{1}^{00}]\mid \lambda ,\mu \rangle =k(k-1)\mid \lambda ,\mu \rangle
\;\;\;,\;\;\;\chi _{1}^{01}\mid \lambda ,\mu \rangle =\mu \mid \lambda ,\mu
\rangle,
\end{equation}%
where $k=\frac{1}{2}+\imath \lambda ,\mu =0,\pm 1,\pm 2,\ldots $, while $%
\lambda $ is an arbitrary real number (see below).

The theory of representations of the SO(2,1)-algebra is a very well
developed area of theoretical physics (see, e.g., \cite{Perl} and references
therein). The non-compact SO(2,1)-algebra and its representations were
introduced for the first time by Bargmann \cite{Barg}. The most detailed and
complete analysis of the SO(2,1)-algebra and its representations can be
found in \cite{Gelf}. Nevertheless, in applications to the problems of 2DG
the Casimir operator $\hat{C}_{2}$ of the SO(2,1)-algebra plays a very
restricted, secondary role. The actual Hamiltonian of the problem $H_{c}$
has a different structure. Below, the structure of $H_{c}$ is considered in
detail. One needs to develop an approach which allows one to construct the
eigenvectors of the $H_{c}$ operator from the $\mid k,m\rangle $ basis
vectors (or $\mid \lambda ,\mu \rangle $ basis vectors).

\section{Properties of the $H_{c}$ Hamiltonian}

The most important feature of the $H_{c}$ operator (or $H_{c}$ Hamiltonian)
follows from its explicit form, equations (\ref{hc}) and (\ref{hc1}).
Briefly, the $H_{c}$ operator is the linear and homogeneous combination of
the three operators $\chi _{1}^{11}$, $\chi _{1}^{01}$, $\chi _{1}^{00}$ (or 
$K_{-},K_{0},K_{+}$) and three values $\xi _{00}^{1}$, $\xi _{01}^{1}$, $\xi
_{11}^{1}$ which depend upon the components of the metric tensor $g_{\mu \nu
}$ and their derivatives. (Note that the values $\xi _{00}^{1}$, $\xi
_{01}^{1}$, $\xi _{11}^{1}$ are linearly related to the affine connection $%
\Gamma _{\alpha \beta }^{\lambda }$ as 
\begin{equation}
\Gamma _{\alpha \beta }^{\lambda }=\xi _{\alpha \beta }^{\lambda }-\frac{1}{%
D-1}\left( \delta _{\alpha }^{\lambda }\xi _{\beta \sigma }^{\sigma }+\delta
_{\beta }^{\lambda }\xi _{\alpha \sigma }^{\sigma }\right)
\end{equation}%
which can be found by inverting equation (\ref{eqn4})).

The linear invertible relation between $\xi _{\alpha \beta }^{\lambda }$ and 
$\Gamma _{\alpha \beta }^{\lambda }$ means that $\xi _{11}^{1}$, $\xi
_{01}^{1}$ and $\xi _{00}^{1}$ do not form a two-dimensional tensor as $%
\Gamma _{\alpha \beta }^{\lambda }$ is not a tensor. Furthermore, all these
three equal zero identically in the case of a Galilean two-dimensional
system. It is easy to understand that only such values can be used in
General Relativity in order to obey the fundamental principle of
equivalence. This explains why only the Hamiltonian, (\ref{hc}), is
acceptable in GR, but all alternative `Hamiltonians' which can be
constructed from the SO(2,1)-algebra, e.g., taking the Hamiltonian to be $%
H=K_{0}^{2}-K_{0}-K_{+}K_{-}$, make no sense in GR. Below, to emphasize the
non-tensorial nature of $\xi _{11}^{1}$, $\xi _{01}^{1}$ and $\xi _{00}^{1}$
we shall call them the $\xi -$symbols. Note that in this notation the affine
connections $\Gamma _{\alpha \beta }^{\lambda }$ can be also recognized as
the $\xi -$symbols.

The Hamiltonian, $H_{c}$, can be presented in the equivalent form as 
\begin{equation}
H_{c}=-2\xi _{01}^{1}K_{0}-\imath (\xi _{11}^{1}-\xi _{00}^{1})K_{1}+(\xi
_{11}^{1}+\xi _{00}^{1})K_{2}=2\sqrt{-G}(n_{0}\cdot K_{0}-n_{1}\cdot
K_{1}-n_{2}\cdot K_{2}),  \label{hc2}
\end{equation}%
where $G=\xi _{00}^{1}\xi _{11}^{1}-(\xi _{01}^{1})^{2}$ is the determinant
of the symmetric $2\times 2$ matrix formed from three $\xi -$numbers: $\xi
_{00}^{1}$, $\xi _{01}^{1},$ and $\xi _{11}^{1}$. We have introduced the
three-dimensional unit-norm vector $\mathbf{n}=(n_{0},n_{1},n_{2})$ (in
pseudo-Euclidean $(2,1)$-space) whose components are 
\begin{equation}
n_{0}=-\frac{\xi _{01}^{1}}{\sqrt{-G}}\;\;\;,\;\;\;n_{1}=\imath \frac{(\xi
_{11}^{1}-\xi _{00}^{1})}{2\sqrt{-G}}\;\;\;,\;\;\;n_{2}=-\frac{(\xi
_{11}^{1}+\xi _{00}^{1})}{2\sqrt{-G}}.  \label{compon}
\end{equation}%
For this vector $\mathbf{n}$ we always have $n_{0}^{2}-n_{1}^{2}-n_{2}^{2}=1$%
. All vectors used below are in the pseudo-euclidean $(2,1)$-space only.
These vectors in the $(2,1)$ pseudo-euclidean space have nothing to do with
the real vectors and/or tensors in the original two-dimensional Riemannian
spacetime with the metric $g_{\mu \nu }$. The actual vectors and/or tensors
transform according the rules dictated by the metric tensor which has three
independent components $g_{00},g_{01}(=g_{10})$ and $g_{11}$ in
two-dimensional spacetime. In contrast, vectors in the $(2,1)$
pseudo-euclidean space are only formal constructions. They are needed to
describe self-adjoint, irreducible representations of the non-compact
SO(2,1)-algebra.

The general form of the $H_{c}$ Hamiltonian of (\ref{hc1}), $H_{c}\simeq (%
\mathbf{n}\cdot \mathbf{K})$, is very similar to the chirality operator $(%
\mathbf{n}\cdot \mathbf{S})$ for moving particles where $\mathbf{S}$ is the
spin and $\mathbf{n}$ is the direction of motion. Analogous chirality
operators are defined for various fields. In the case of the 2DG model the
Hamiltonian, $H_{c}$ in equation (\ref{hc1}), is formally written as the
scalar product of the two $(2,1)$-vectors $\mathbf{K}=(K_{0},K_{1},K_{2})$
and $\mathbf{n}=(n_{0},n_{1},n_{2})$. The vector $\mathbf{n}$ can be
considered as a `direction of propagation' of the free field defined in the $%
(2,1)$ pseudo-euclidean space. All components of this vector are $\xi -$%
symbols. Analogically the vector $\mathbf{K}$ represents some internal
property of this field. The components of this $(2,1)$-vector, $\mathbf{K}$,
are the three secondary constraints of 2DG. All secondary constrains do not
change with time (see, e.g., \cite{Dir}), so in this sense the Hamiltonian, $%
H_{c}$, can be considered as the `chirality operator' of the 2DG model.

\section{Eigenvectors of the $H_{c}$ Hamiltonian}

The Hamiltonian, $H_{c}$, in (\ref{hc2}) is the linear form of the three $%
\xi -$numbers and the three generators of the SO(2,1)-algebra. In general,
for an arbitrary operator which is represented as a linear combination of
three generators of this algebra, there is a well developed procedure which
can be used to obtain the eigenvalues and eigenvectors of this operator
based on the use of coherent states of the SO(2,1)-algebra.

The system of coherent states for the discrete series of SO(2,1)-algebra
representations has been constructed in \cite{Perl}. In our paper we shall
follow the procedure described in \cite{Perl}. At the first step one chooses
an arbitrary vector $\mid \psi _{0}\rangle $. It is shown in \cite{Perl}
that there are some advantages to choosing such a vector to be in the form $%
\mid \psi _{0}\rangle =\mid k,k\rangle $, i.e. the vector $\mid k,k+m\rangle 
$ for which $m=0$. The coherent states derived from the $\mid k,k\rangle $
vector have properties which are similar to the properties of the
corresponding `classical' states. At the second stage we represent the
unit-norm pseudo-euclidean vector $\mathbf{n}=(n_{0},n_{1},n_{2})$ in the
following two-parameter form

\begin{equation}
(n_{0},n_{1},n_{2})=(\cosh \tau ,\sinh \tau \cos \phi ,\sinh \tau \sin \phi).
\label{n}
\end{equation}

These vectors can be used to designate a corresponding coherent state $\mid 
\mathbf{n}\rangle $. Moreover, $\mid \mathbf{n}\rangle =D(\mathbf{n})\mid
\psi _{0}\rangle $, where $D(\mathbf{n})$ is some operator which is
represented in the following three-parameter form 
\begin{equation}
D(\mathbf{n})=\exp (\alpha K_{-})\exp (\beta K_{0})\exp (\gamma K_{+}),
\label{opD}
\end{equation}%
where $\beta =-\ln (1-\mid \alpha \mid ^{2})$ and $\gamma =-\overline{\alpha 
}$ ($\overline{z}$ is the complex conjugate to $z$). The coherent state $%
\mid \mathbf{n}\rangle (=D(\mathbf{n})\mid \psi _{0}\rangle $) can be
designated with the use of this one parameter ($\alpha $) only (see,
equation (\ref{e19}) below). There is a relation between the parameter $%
\alpha $ and three components of the vector $\mathbf{n}$ (\ref{n}), given by 
$\alpha =\tanh (\frac{\tau }{2})\exp (\imath \phi )$. The transition from
variables $\mathbf{n}$ (or $\tau ,\phi $) to the variable $\alpha $
corresponds to the stereographic projection from south pole of hyperboloid,
i.e., $\mathbf{n}_{0}=(-1,0,0)$, on the complex $\alpha $-plane.

The operator $D(\mathbf{n})$ defined in equation (\ref{opD}) is similar to
the $D$-matrix known for the compact $SO(3)$-algebra. Here we do not discuss
this analogy in detail (such a discussion can be found in \cite{Perl}, see
also references therein). For our present analysis it is important to write
the coherent states as infinite expansions upon the basis set of unit-norm
vectors $\mid k,m\rangle $ defined above for the positive series of
representations (see equations (\ref{basic}) and (\ref{basic1})) 
\begin{equation}
\mid \alpha \rangle =\mid \alpha ,\beta (\alpha ),\gamma (\alpha )\rangle
=(1-\mid \alpha \mid ^{2})\sum_{\infty }^{m=0}\sqrt{\frac{(m+2k-1)!}{%
m!(2k-1)!}}\cdot \alpha ^{m}\mid k,k+m\rangle ,  \label{e19}
\end{equation}%
where $\alpha $ is one of the three parameters of the coherent state $\mid
\alpha ,\beta ,\gamma \rangle $. The properties of these coherent states are
discussed in \cite{Perl}. The most important of these properties is: \textit{%
this state is an eigenstate of the} $\left( n_{0}\cdot K_{0}-n_{1}\cdot
K_{1}-n_{2}\cdot K_{2}\right) $ \textit{operator}. This immediately follows
from our choice of the unit vector $\mid \mathbf{n}\rangle $ and from the
identity $D(\mathbf{n})K_{0}D^{-1}(\mathbf{n})=(\mathbf{n}\cdot \mathbf{K})$%
, where $\mathbf{K}=(K_{0},K_{1},K_{2})$. Indeed, from the definition $\mid 
\mathbf{n}\rangle =D(\mathbf{n})\mid \psi _{0}\rangle $ we can write $D^{-1}(%
\mathbf{n})\mid \mathbf{n}\rangle =\mid \psi _{0}\rangle $; and therefore 
\begin{equation}
K_{0}(D^{-1}(\mathbf{n})\mid \mathbf{n}\rangle )=K_{0}\mid \psi _{0}\rangle
=k\mid \psi _{0}\rangle ,
\end{equation}%
since $\mid \psi _{0}\rangle $ was chosen to be an eigenvector of $K_{0}$.
From here one finds 
\begin{equation}
D(\mathbf{n})K_{0}D^{-1}(\mathbf{n})\mid \mathbf{n}\rangle =kD(\mathbf{n}%
)\mid \psi _{0}\rangle =k\mid \mathbf{n}\rangle .  \label{e21}
\end{equation}%
On the other hand, we have the identity $D(\mathbf{n})K_{0}D^{-1}(\mathbf{n}%
)=(\mathbf{n}\cdot \mathbf{K})$. By combining equation (\ref{e21}) and this
identity we obtain 
\begin{equation}
(\mathbf{n}\cdot \mathbf{K})\mid \mathbf{n}\rangle =k\mid \mathbf{n}\rangle .
\label{e22}
\end{equation}%
In other words, the vector $\mid \mathbf{n}\rangle $ is an eigenvector of
the $(\mathbf{n}\cdot \mathbf{K})$ operator and $k$ is its eigenvalue.

In our notation this eigenvalue equation can also be written in the form $(%
\mathbf{n}\cdot \mathbf{K})\mid \alpha \rangle =k\mid \alpha \rangle $,
where $\alpha $ is the complex parameter which determines the coherent state
for the discrete series of representations of the SO(2,1)-algebra. It
follows from here that for the $\mid \alpha \rangle $ vector the following
equation is also obeyed 
\begin{equation}
H_{c}\mid \alpha \rangle =2\sqrt{-G}(n_{0}\cdot K_{0}-n_{1}\cdot
K_{1}-n_{2}\cdot K_{2})\mid \alpha \rangle =2k\sqrt{-G}\mid \alpha \rangle .
\label{hc4}
\end{equation}%
This means that the coherent state $\mid \alpha \rangle (=\mid \alpha ,\beta
,\gamma \rangle $ is the eigenvector of the $H_{c}$ operator with the
eigenvalue $\lambda =2k\sqrt{-G}$. This eigenvalue equals zero in any flat
two-dimensional spacetime. It should be mentioned that there is another
condition which is always obeyed for the $\mid \alpha \rangle $ vector \cite%
{Perl}, 
\begin{equation}
(K_{-}-2\alpha K_{0}+\alpha ^{2}K_{+})\mid \alpha \rangle =0.
\end{equation}%
The role of this condition for the 2DG model is not quite clear, since it
contains a mixture of the regular numbers (unity) and $\xi -$values ($\alpha 
$ and $\alpha ^{2}$).

The discrete series of self-adjoint irreducible representations of the $%
SO(2,1)-$algebra constructed above (see equations (\ref{e19}) and (\ref{e22}%
)) is quite restricted when considering actual problems of GR. It is clear
that we need to construct analogous coherent states for the main (or
continuous) series of self-adjoint irreducible representations of the
SO(2,1)-algebra. Let $\mid m\rangle \equiv \mid k,m\rangle $ be the
unit-norm basis in some Hilbert space $\mathcal{H}$. All these vectors are
the eigenvectors of the operator $K_{0}$, i.e. $K_{0}\mid \mu \rangle =\mu
\mid \mu \rangle $. Also, they are eigenvectors of the Casimir operator $%
\hat{C}_{2}$ defined above. The coherent states can be expanded in terms of
these basis vectors. Below, we shall designate the corresponding coherent
state by $\mid \alpha \rangle $, while the notation $\mid m\rangle $ always
means the eigenvectors of the $K_{0}$ and $\hat{C}_{2}$ operators.

The coherent state $\mid \alpha \rangle $ can be represented as an infinite
sum of eigenstates $\mid m\rangle $, i.e. 
\begin{equation}
\mid \alpha \rangle =\sum_{n=-\infty }^{\infty }u_{n}(\alpha )\mid n\rangle
\end{equation}%
where the coefficients $u_{n}(\alpha )$ \cite{Perl} are 
\begin{equation}
u_{n}(\alpha )=\langle n\mid \alpha \rangle =u_{n}^{\lambda }(\tau ,\phi ).
\label{e25}
\end{equation}%
Here $\tau $ and $\phi $ are the parameters which define the unit-norm
pseudo-euclidean vector $\mathbf{n}=(n_{0},n_{1},n_{2})$ (\ref{n}). For
continuous series of the irreducible representation of the SO(2,1)-algebra
the relation between parameters $\alpha $ and $\tau ,\phi $ is $\alpha
=-\tanh (\frac{\tau }{2})\exp (\imath \phi )$; i.e. it differs by sign from
the analogous relation used above for the discrete series.

The coefficients $u_{n}^{\lambda }(\tau ,\phi )$ defined in equation(\ref%
{e25}) have three following properties \cite{Perl} 
\begin{equation}
u_{n}^{\lambda }(0,\phi )=\delta _{0n}\ ,\;\;\;u_{n}^{\lambda }(\tau ,\phi
)=\exp (-\imath n\phi )R_{n}^{\lambda }(\tau )  \label{Un1}
\end{equation}%
and 
\begin{equation}
\sum_{n=-\infty }^{n=\infty }\mid u_{n}^{\lambda }(\tau ,\phi )\mid ^{2}=1
\end{equation}%
for arbitrary $\tau ,\phi $ and $\lambda $. The last equality follows from
the fact that all coherent states $\mid \alpha \rangle $ have unit norm.

It can also be shown that these coefficients $u_{n}^{\lambda }(\tau ,\phi )$
coincide with the corresponding eigenfunctions of the Laplace-Beltrami
operator $\tilde{\Delta}$ (see, e.g., \cite{Flan}) constructed for the
Lobachevskii plane; i.e. 
\begin{equation}
\tilde{\Delta}u_{n}^{\lambda }(\tau ,\phi )=\Bigl[\frac{\partial ^{2}}{%
\partial \tau ^{2}}+\cosh \tau \frac{\partial }{\partial \tau }+\frac{1}{%
\sinh ^{2}\tau }\frac{\partial ^{2}}{\partial \phi ^{2}}\Bigr]u_{n}^{\lambda
}(\tau ,\phi )=\Lambda u_{n}^{\lambda }(\tau ,\phi )=-\left( \frac{1}{4}%
+\lambda ^{2}\right) u_{n}^{\lambda }(\tau ,\phi )  \label{Beltr}
\end{equation}%
for principal series of SO(2,1)-representations $\lambda $ is an arbitrary
real number, while for the supplementary series: $\lambda =\imath \sigma $,
where $\sigma $ is also real, but $\mid \sigma \mid \leq \frac{1}{2}$.
Below, we shall consider only the principal series.

Note that the $u_{n}^{\lambda }(\tau ,\phi )$ functions are the
eigenfunctions of the two commuting and self-adjoint operators $\tilde{\Delta%
}$ and $-\imath \frac{\partial }{\partial \phi }$. This means that these
functions of the $\tau $ and $\phi $ variables form a complete system of
orthogonal (basis) functions on the Lobachevskii plane. In other words, an
arbitrary function of the $\tau $ and $\phi $ variables can be approximated
by linear combinations of the $u_{n}^{\lambda }(\tau ,\phi )$ functions. For
the principal series of representations of the SO(2,1)-algebra the
orthogonality relation for the $u_{n}^{\lambda }(\tau ,\phi )$ functions
takes the form 
\begin{equation}
\frac{1}{2\pi }\int_{0}^{\infty }\int_{0}^{2\pi }\overline{u}_{n}^{\lambda
}(\tau ,\phi )u_{n_{1}}^{\lambda _{1}}(\tau ,\phi )\sinh \tau d\tau d\phi
=N_{n}(\lambda )\delta _{nn_{1}}\delta (\lambda -\lambda _{1}).
\end{equation}

To conclude this section we want to note that there is an obvious analogy
between the $u_{n}^{\lambda }(\tau ,\phi )$ functions in the Lobachevskii
plane and the plane waves $\exp (\imath \mathbf{k}\mathbf{r})$ in the
Euclidean plane (for more details, see \cite{Perl}). The second comment is
related to the fact that the functions $\sqrt{\sinh \tau }R_{n}^{\lambda
}(\tau )$, where $R_{n}^{\lambda }(\tau )$ are defined in equation (\ref{Un1}%
), obey the non-relativistic Schr\"{o}dinger equation with the (scattering)
potential $V(\tau )=(n^{2}-\frac{1}{4})\sinh ^{-2}\tau $. This analogy
allows one to obtain many additional properties of the $u_{n}^{\lambda
}(\tau ,\phi )$ functions.

\section{The Hamilton-Jacobi method}

It is crucial for our analysis that the canonical Hamiltonian density of the
problem, $H_{c}$ (\ref{hc}), contains only secondary constraints which do
not change with time $t$ \cite{Dir}. The only variables which are included
in the $H_{c}$ Hamiltonian density are $\xi _{00}^{1}$, $\xi _{01}^{1}$ and $%
\xi _{11}^{1}$. The variables ($t$, $\xi _{00}^{1}$, $\xi _{01}^{1}$ and $%
\xi _{11}^{1}$) can be considered as the four actual variables of the
problem. In this case it follows from (\ref{hc}) that 
\begin{equation}
H_{c}\delta t=-\chi _{1}^{11}\delta \xi _{11}^{1}-2\chi _{1}^{01}\delta \xi
_{01}^{1}-\chi _{1}^{00}\delta \xi _{00}^{1}  \label{hcdelt}
\end{equation}%
or, in other words, 
\begin{equation}
\frac{\partial H_{c}}{\partial \xi _{00}^{1}}=-\chi _{1}^{11}\;\;\;,\;\;\;%
\frac{\partial H_{c}}{\partial \xi _{01}^{1}}=-2\chi _{1}^{01}\;\;\;,\;\;\;%
\frac{\partial H_{c}}{\partial \xi _{00}^{1}}=-\chi _{1}^{00}.
\end{equation}%
This means that we can write 
\begin{equation}
H_{c}=\xi _{11}^{1}\cdot \frac{\partial H_{c}}{\partial \xi _{11}^{1}}+\xi
_{01}^{1}\cdot \frac{\partial H_{c}}{\partial \xi _{01}^{1}}+\xi
_{00}^{1}\cdot \frac{\partial H_{c}}{\partial \xi _{00}^{1}}
\end{equation}%
where all partial derivatives on the right-hand side do not depend upon $t$.
Bearing this equation in mind, let us try to find the function $S(t,\xi
_{00}^{1},\xi _{01}^{1},\xi _{11}^{1})$ for which the following equation is
always obeyed 
\begin{equation}
\frac{\partial S}{\partial t}=-\xi _{11}^{1}\frac{\partial S}{\partial \xi
_{11}^{1}}-\xi _{01}^{1}\frac{\partial S}{\partial \xi _{01}^{1}}-\xi
_{00}^{1}\frac{\partial S}{\partial \xi _{00}^{1}}.  \label{e5}
\end{equation}%
In particular, we can try to represent the function $S(t,\xi _{00}^{1},\xi
_{01}^{1},\xi _{11}^{1})$ in the form $S(t,\xi _{00}^{1},\xi _{01}^{1},\xi
_{11}^{1})=f(t)\cdot S_{0}(\xi _{00}^{1},\xi _{01}^{1},\xi _{11}^{1})$,
where $S_{0}(x,y,z)$ is a homogeneous function of power $b$, i.e. 
\begin{equation}
x\frac{\partial S_{0}}{\partial x}+y\frac{\partial S_{0}}{\partial y}+z\frac{%
\partial S_{0}}{\partial z}=bS_{0},
\end{equation}%
where $b$ is a real number. In this case from (\ref{e5}) one finds 
\begin{equation}
\frac{df(t)}{dt}=-bf(t),
\end{equation}%
or $f(t)=A\exp (-bt)$. Therefore, the function $S(t,\xi _{00}^{1},\xi
_{01}^{1},\xi _{11}^{1})$ does exist and can be found. Moreover, it can be
presented in the form 
\begin{equation}
S(t,\xi _{00}^{1},\xi _{01}^{1},\xi _{11}^{1})=A\exp (-bt)\cdot S_{0}(\xi
_{00}^{1},\xi _{01}^{1},\xi _{11}^{1}),
\end{equation}%
where $S_{0}(x,y,z)$ is an arbitrary homogeneous function of power $b$. In
the last equation the variables $\xi _{00}^{1},\xi _{01}^{1}$ and $\xi
_{11}^{1}$ do not depend upon $t$. The function $S(t,\xi _{00}^{1},\xi
_{01}^{1},\xi _{11}^{1})$ depends upon time $t$ only by the exponential
factor (decaying factor for $b>0$). In general, the function $S(t,\xi
_{00}^{1},\xi _{01}^{1},\xi _{11}^{1})$ is the Jacobi function, while $S_{0}$
is the so-called short Jacobi function.

\section{Second quantized form of the $H_{c}$ Hamiltonian}

The Hamiltonian, $H_{c}$, can be represented in a second quantized form. In
fact, such a form immediately follows from equation (\ref{hc2}) and the
following theorem about unitary representations of the SO(2,1)-algebra. Let $%
a_{1},a_{2},a_{1}^{+}$ and $a_{2}^{+}$ be the four bosonic operators for
which the following commutation relations are obeyed $[a_{i},a_{j}^{+}]=%
\delta _{ij},[a_{i}^{+},a_{j}^{+}]=0$ and $[a_{i},a_{j}]=0$, where $i$ = 1,
2 and $j$ = 1, 2. In this case the three following operators 
\begin{equation}
X_{1}=\frac{\imath }{2}(a_{1}^{+}a_{2}+a_{2}^{+}a_{1})\;\;\;,\;\;\;X_{2}=-%
\frac{1}{2}(a_{1}^{+}a_{2}-a_{2}^{+}a_{1})\;\;\;,\;\;\;X_{0}=\frac{1}{2}%
(a_{1}^{+}a_{1}-a_{2}^{+}a_{2})  \label{Ap1}
\end{equation}%
form the SO(2,1)-algebra \cite{Bar}. Furthermore, let $\mid \phi ,m\rangle
=N_{\phi ,m}(a_{1}^{+})^{\phi +m}(a_{2}^{+})^{\phi -m}\mid 0\rangle $ be the
basis vectors, where $a_{1}\mid 0\rangle =0$ and $a_{2}\mid 0\rangle =0$. On
these vectors one finds for the operators $X_{0}$ and $\hat{C}%
_{2}=X_{0}^{2}-X_{1}^{2}-X_{2}^{2}$ 
\begin{equation}
X_{0}\mid \phi ,m\rangle =m\mid \phi ,m\rangle \;\;\;,\;\;\;\hat{C}_{2}\mid
\phi ,m\rangle =\phi (\phi +1)\mid \phi ,m\rangle .
\end{equation}%
The operator $\hat{C}_{2}$ is the Casimir operator of the SO(2,1)-algebra.

Let us discuss the unitary representations of the SO(2,1)-algebra. In this
case $\phi (\phi +1)$ must be real and all three operators $%
X_{0},X_{1},X_{2} $ must be self-adjoint. The positive series of unitary
representations of the SO(2,1)-algebra is obtained by applying the condition 
$m=-\phi ,-\phi +1,\ldots ,-\phi +n,\ldots $. The negative series
corresponds to the choice $m=\phi ,\phi -1,\ldots ,\phi -n,\ldots $. The
choice $\phi =-\frac{1}{2}+\imath \rho $, where $\rho $ is real, represents
the the principal series of unitary representations of the SO(2,1)-algebra.
Here we do not want to discuss applications of this theorem to various
problems. Note, however, that this theorem allows one to represent the
Hamiltonian, $H_{c}$, in the second quantized form. Indeed, from equation (%
\ref{hc2}) and equation (\ref{Ap1}) one finds

\begin{eqnarray}
H_{c} &=&\sqrt{-G}\Bigl[n_{0}\cdot (a_{1}^{+}a_{1}-a_{2}^{+}a_{2})-\imath
\cdot n_{1}\cdot (a_{1}^{+}a_{2}+a_{2}^{+}a_{1})-n_{2}\cdot
(a_{1}^{+}a_{2}-a_{2}^{+}a_{1})\Bigr]  \label{51} \\
&=&-\xi _{01}^{1}(a_{1}^{+}a_{1}-a_{2}^{+}a_{2})+\xi
_{11}^{1}a_{1}^{+}a_{2}-\xi _{00}^{1}a_{2}^{+}a_{1},  \notag
\end{eqnarray}
where all parameters in this formula have been defined in the main text.
This form of $H_{c}$ is of interest in Quantum Gravity.

Note that the commutation relations between operators $a_{i}$ and $a_{j}^{+}$
($i,j$ = 1, 2) mentioned above exactly coincide with the Poisson brackets
between a set of canonical variables known from the Hamilton Classical
Mechanics (see, e.g., \cite{LLM}). Therefore, we can consider, Eq.(\ref{hc}%
), as the Hamiltonian $H_{c}$ which is already written in canonical
variables. Let us obtain the canonical equations for four operators $a_{i}$
and $a_{j}^{+}$ ($i,j$ = 1, 2), i.e. for these canonical variables. By using
the explicit form of $H_{c}$, Eq.(\ref{hc}), one finds 
\begin{eqnarray}
\frac{da_{1}}{dt} &=&[a_{1},H_{c}]=-\xi _{01}^{1}a_{1}+\xi _{11}^{1}a_{2},
\label{em1-1} \\
\frac{da_{2}}{dt} &=&[a_{2},H_{c}]=-\xi _{00}^{1}a_{1}+\xi _{01}^{1}a_{2}.
\label{em2}
\end{eqnarray}%
Analogous equations for the $a_{j}^{+}$ ($j$ = 1, 2) are 
\begin{eqnarray}
\frac{da_{1}^{+}}{dt} &=&[a_{1}^{+},H_{c}]=\xi _{01}^{1}a_{1}^{+}+\xi
_{00}^{1}a_{2}^{+},  \label{em3} \\
\frac{da_{2}^{+}}{dt} &=&[a_{2}^{+},H_{c}]=-\xi _{11}^{1}a_{1}^{+}-\xi
_{01}^{1}a_{2}^{+}.  \label{em4}
\end{eqnarray}%
To determine the actual time-dependence of these operators, let us assume
that $a_{i}(t)=a_{i}(0)\exp (\imath \omega t)$, and therefore, $%
a_{i}^{+}(t)=a_{i}^{+}(0)\exp (-\imath \omega t)$. In this case from
equations of motion Eqs.(\ref{em1-1}) - (\ref{em2}) one obtains 
\begin{eqnarray}
\imath \omega a_{1} &=&-\xi _{01}^{1}a_{1}+\xi _{11}^{1}a_{2},  \label{em1a}
\\
\imath \omega a_{2} &=&-\xi _{00}^{1}a_{1}+\xi _{01}^{1}a_{2}.  \label{em2a}
\end{eqnarray}%
This system of equations has a non-trivial solution if the determinant of
the following $2\times 2$ matrix 
\begin{equation}
\left( 
\begin{array}{cc}
\xi _{01}^{1}+\imath \omega & -\xi _{11}^{1} \\ 
\xi _{00}^{1} & -\xi _{01}^{1}+\imath \omega%
\end{array}%
\right)  \notag
\end{equation}%
equals zero. In this case the solutions are 
\begin{equation}
\omega _{1,2}=\pm \sqrt{(\xi _{01}^{1})^{2}-\xi _{00}^{1}\xi _{11}^{1}}.
\end{equation}%
Without loss of generality, we shall choose the positive root, i.e. $\omega
=\omega _{1}=\sqrt{(\xi _{01}^{1})^{2}-\xi _{00}^{1}\xi _{11}^{1}}$. In this
case the evolution in time of the $a_{1}(t)$ and $a_{2}(t)$ operators is
represented in the form $a_{i}(t)=a_{i}(0)\exp (\imath \omega t)$. Other
possible forms of time dependence for these two operators will not be
discussed in this study. Now, we can introduce the two pairs of conjugate
operators $Q_{i}$ and $P_{i}$ which are simply and canonically related to
the $a_{i}$ and $a_{i}^{+}$ operators 
\begin{equation}
a_{i}=\frac{1}{\sqrt{2\omega }}(\omega Q_{i}+\imath P_{i})  \label{ec1}
\end{equation}%
and 
\begin{equation}
a_{i}^{+}=\frac{1}{\sqrt{2\omega }}(\omega Q_{i}-\imath P_{i}).  \label{ec2}
\end{equation}%
The inverse relations take the form 
\begin{equation}
Q_{i}=\frac{1}{\sqrt{2\omega }}(a_{i}+a_{i}^{+})  \label{ec3}
\end{equation}%
and 
\begin{equation}
P_{i}=\imath \sqrt{\frac{\omega }{2}}(a_{i}^{+}-a_{i}).  \label{ec4}
\end{equation}%
Note that the operators $P_{i}$ are Hermitian. From (\ref{ec3}) and (\ref%
{ec4}) one finds 
\begin{equation}
\lbrack Q_{i},Q_{j}]=0\;\;\;,\;\;\;\;[Q_{i},P_{j}]=\imath \delta
_{ij}\;\;(or\;\;[P_{i},Q_{j}]=-\imath \delta
_{ij})\;\;\;,\;\;\;[P_{i},P_{j}]=0  \label{PB1}
\end{equation}%
where $i$ = 1, 2 and $j$ = 1, 2. In other words, the two `coordinates' $%
Q_{1},Q_{2}$ and two momenta $P_{1},P_{2}$ can be considered as the new
canonical variables, which are related to the old canonical variables $%
a_{1},a_{2},a_{1}^{+},a_{2}^{+}$ by a canonical transformation, Eqs.(\ref%
{ec1}) - (\ref{ec4}). The Hamiltonian $H_{c}$, Eq.(\ref{51}), written in
these new canonical variables takes the form 
\begin{eqnarray}
H_{c} &=&-\ \frac{\xi _{01}^{1}}{2\sqrt{(\xi _{01}^{1})^{2}-\xi _{00}^{1}\xi
_{11}^{1}}}\Bigl[\Bigl(P_{1}^{2}+\omega ^{2}Q_{1}^{2}\Bigr)-\Bigl(%
P_{2}^{2}+\omega ^{2}Q_{2}^{2}\Bigr)\Bigr]  \notag \\
&+&\frac{(\xi _{11}^{1}-\xi _{00}^{1})}{2\sqrt{(\xi _{01}^{1})^{2}-\xi
_{00}^{1}\xi _{11}^{1}}}\Bigl(P_{1}P_{2}+\omega ^{2}Q_{1}Q_{2}\Bigr)
\label{hamq} \\
&-&\frac{\imath (\xi _{11}^{1}+\xi _{00}^{1})}{2\sqrt{(\xi
_{01}^{1})^{2}-\xi _{00}^{1}\xi _{11}^{1}}}(Q_{1}P_{2}-Q_{2}P_{1}).  \notag
\end{eqnarray}%
By using this form of $H_{c}$ written in the canonical variables $%
P_{1},P_{2},Q_{1},Q_{2}$ it is straightforward to derive the corresponding
canonical equations for these variables. In general, the Hamiltonian
density, Eq.(\ref{hamq}), corresponds to the case of two coupled harmonic
oscillators. Formally, the quantization of the canonical equations for the $%
P_{1},P_{2},Q_{1},Q_{2}$ variables does not present any difficulty. Note
that the coupling of the two classical oscillators is described by the two
last terms in Eq.(\ref{hamq}). Briefly, we can say that such a coupling
cannot be found in any vibrational system known in classical mechanics. The
next step of our procedure is to perform the explicit quantization of the
Hamiltonian $H_{c}$ and a related system of canonical equations which
represent the two-dimensional gravity.

\section{Conclusion}

We have analyzed the algebraic structure of the model of two-dimensional
gravity. It is shown that the algebraic structure of this model is locally
isomorphic to the SO(2,1)-algebra. The canonical Hamiltonian of 2DG is
expressed as a linear combination of the three generators of this algebra.
These generators coincide with the three secondary first class constraints
defined in the model. These secondary constraints do not change with time.
The coefficients included in the Hamiltonian are the $\xi -$numbers which
are uniformly related to the affine connections $\Gamma _{\alpha \beta
}^{\lambda }$. The linear form of the Hamiltonian allows us to apply a
special procedure which has been developed earlier to determine its
eigenvalues. This procedure is based on the use of coherent states for the
SO(2,1)-algebra. We consider the coherent states constructed for the
discrete and principal series of representations of the SO(2,1)-algebra.

Finally, the analysis of the 2DG model can be presented in the following
way. If an arbitrary point in the actual two-dimensional metric space $%
g_{00},g_{01},g_{11}$ is given, then we can define the tensor density $%
h^{\alpha \beta }=\sqrt{-g}g^{\alpha \beta }$. By using the components of
this tensor we can determine the momenta, $\pi _{\alpha \beta }$, conjugate
to each of the components. One finds an explicit formula for the three
secondary constraints $\chi _{1}^{11},\chi _{1}^{01}$ and $\chi _{1}^{00}$.
The PB algebra of the constraints is an SO(2,1)-algebra, while the canonical
Hamiltonian of 2DG is written as a linear combination of these three
secondary first class constraints.

It should be mentioned that coherent states for non-Abelian SO(3)-algebra
have been constructed for the first time by Radcliffe in \cite{Radc}.
Perelomov \cite{Per2} considered a more general case of arbitrary
non-Abelian algebras. The method developed in \cite{Per2} (see also \cite%
{Perl}) allows one to construct various systems of coherent states for many
non-Abelian algebras, including SO(2,1), SO(3,1), SO(N,1) algebras and
others. The importance of coherent states in our analysis is based on the
fact that such states essentially coincide with the corresponding
eigenvectors of the canonical Hamiltonian $H_{c}$. In turn, the $H_{c}$
Hamiltonian plays a central role in the 2DG model.

As we mentioned above, the model of two-dimensional gravity that we have
considered has a number of advantages. In particular, the methods developed
for our model can be applied to formulations of both metric and tetrad
gravity. For instance, let us consider the Hamiltonian of tetrad gravity
which is derived from its first-order formulation. It was shown in \cite%
{arxiv} that up to a total spatial derivative, the canonical Hamiltonian
density of three-dimensional tetrad gravity is a linear combination of the
secondary first class constraints (\textquotedblleft
rotational\textquotedblright\ $\left( \chi ^{0(\rho )}\right) $ and
\textquotedblleft translational\textquotedblright\ $\left( \chi ^{0(\alpha
\beta )}\right) $ constraints), i.e. 
\begin{equation}
H_{c}=-e_{0(\rho )}\chi ^{0(\rho )}-\omega _{0(\alpha \beta )}\chi
^{0(\alpha \beta )}  \label{eX1}
\end{equation}%
where $e_{0(\rho )}$ are tetrads, while $\omega _{0(\alpha \beta )}$ are the
spin connections. (For notation and the explicit form of the constraints see 
\cite{arxiv}.) The result of \cite{arxiv} allows us to infer that the form
of (\ref{eX1}) is common for all dimensions $D>2$. Further developments
along this line can be found in \cite{Report, Transl} where (\ref{eX1}) is
obtained in higher dimensions. Translational and rotational invariance in
the tangent space is the general property of the first-order tetrad (or
N-bein) gravity in all dimensions $D>2.$ In higher dimensions the only
possible modification is the Poisson bracket among translational constraints
which might differ from zero but proportional to the constraints and in $3D$
limit gives zero \cite{Report, Transl}. This work is in progress and the
results will be reported elsewhere.

The $D$ translational $\chi ^{0(\rho )}$ and $\frac{D(D-1)}{2}$ rotational $%
\chi ^{0(\alpha \beta )}$ constraints of the Poincar\'{e} algebra form a
closed and local algebraic structure. To classify this algebra, let us
redefine the above constraints $M^{\mu \nu }=\frac{1}{2}\chi ^{0(\mu \nu )}$
and $P^{\mu }=\frac{1}{2}\chi ^{0(\mu )}$. In this notation, the Poisson
brackets obtained in \cite{arxiv} for $D=3$ have a $D-$dimensional form 
\begin{eqnarray}
\{M^{\mu \nu },M^{\rho \sigma }\} &=&\eta ^{\nu \rho }M^{\mu \sigma }+\eta
^{\nu \sigma }M^{\mu \rho }-\eta ^{\mu \rho }M^{\nu \sigma }-\eta ^{\mu
\sigma }M^{\nu \rho }\;\;\;,  \label{eX3} \\
\{M^{\mu \nu },P^{\sigma }\} &=&\eta ^{\nu \sigma }P^{\mu }-\eta ^{\mu
\sigma }P^{\nu }\;\;\;,\;\;\;\{P^{\mu },P^{\nu }\}=0  \label{eX5}
\end{eqnarray}%
where $M^{\mu \nu }=-M^{\nu \mu },\mu ,\nu ,\ldots =0,1,2,\ldots $, and $%
\eta ^{\mu \nu }=diag\left( 1,-1,-1,...\right) $ is the Minkowski tensor. In
three-dimensional spacetime the maximal value of indices in these
definitions equals 2. With this notation we see that the PBs, equations (\ref%
{eX3}) - (\ref{eX5}), coincide with the commutation relations known for the
generators of Poincar\'{e} algebra ISO($D-1$,1) = P(1,$D-1$). In $D$
dimensions the Poincar\'{e} algebra ISO($D-1$,1) is represented as a
semi-direct sum of its ideal $t^{D}$ (which contains $D$ translations only)
and corresponding Lorentz group SO($D-1$,1) (which contains the $\frac{D(D-1)%
}{2}$ rotations). So that ISO($D-1$,1) = $t^{D}\Join $ SO($D-1$,1).

With this notation, equation (\ref{eX1}) gives a proportional to the
secondary first class constraints part of the canonical Hamiltonian density
which takes the form 
\begin{equation}
H_{c}=-2e_{0(\rho )}P^{\rho }-2\omega _{0(\alpha \beta )}M^{\alpha \beta }.
\label{eX11}
\end{equation}%
As follows from this equation, this part is linear in the secondary first
class constraints $P^{\rho }$ and $M^{\alpha \beta }$ and also linear in the
tetrads $e_{0(\rho )}$ and spin connections $\omega _{0(\alpha \beta )}$.
This form of the Hamiltonian $H_{c}$ is common in all problems related to
tetrad gravity. Since $e_{0(\rho )}$ and $\omega _{0(\alpha \beta )}$ in
equation (\ref{eX11}) are real, while all generators $P^{\rho }$ and $%
M^{\alpha \beta }$ are self-adjoint, all eigenvectors and corresponding
eigenvalues of the Hamiltonian $H_{c}$ can be found with the use of the
procedure which is generalization of our method for the case of Poincar\'{e}
algebra. Formally, all these eigenvectors and eigenvalues will contain $%
e_{0(\rho )}$ and $\omega _{0(\alpha \beta )}$ as parameters. This means
that there is a remarkable analogy of our simple model of two-dimensional
gravity and tetrad gravity for an arbitrary $D-$dimensional spacetime.

\vspace{5mm}

\textbf{Acknowledgements}

We would like to thank P.G. Komorowski and D.G.C. McKeon for helpful
discussions and reading the manuscript.

\end{document}